# Evidence for high-energy and low-emittance electron beams using ionization injection of charge in a plasma wakefield accelerator

N Vafaei-Najafabadi[1], W An[1], C E Clayton[1], C Joshi[1], K A Marsh[1], W B Mori[1], E C Welch[1], W Lu[2], E Adli[3,4], J Allen[3], C I Clarke[3], S Corde[3,5], J Frederico[3], S J Gessner[3], S Z Green[3], M J Hogan[3], M D Litos[3], and V Yakimenko[3]

[1] University of California Los Angeles, Los Angeles, CA 90095, USA
[2] Department of Engineering Physics, Tsinghua University, Beijing 100084, China
[3] SLAC National Accelerator Laboratory, Menlo Park, CA 94025, USA

E-mail: navidvafa@ucla.edu

**Abstract**

Ionization injection in a plasma wakefield accelerator was investigated experimentally using two lithium plasma sources of different lengths. The ionization of the helium gas, used to confine the lithium, injects electrons in the wake. After acceleration, these injected electrons were observed as a distinct group from the drive beam on the energy spectrometer. They typically have a charge of tens of pC, an energy spread of a few GeV, and a maximum energy of up to 30 GeV. The emittance of this group of electrons can be many times smaller than the initial emittance of the drive beam. The energy scaling for the trapped charge from one plasma length to the other is consistent with the blowout theory of the plasma wakefield.

## 1. Introduction

In the future, both collider and light source applications will require extremely low emittance electron beams. We are therefore investigating if it is possible to use the plasma wakefield acceleration (PWFA) scheme to produce such low emittance beams. In particular, we explore the use of a relatively high emittance drive bunch to produce a wake that can accelerate a lower emittance bunch to energies comparable to the drive bunch energy. In this paper, we report our initial findings where we employ the ionization-injection technique to inject the electrons in the wake, which are accelerated at high gradients. This self-generated, secondary bunch contains tens of pC of charge and energies up to 30 GeV. Furthermore, emittance measurements of the injected beam imply an emittance value that is many times smaller than the initial emittance of the drive beam.

---

[4] Currently at Department of Physics, University of Oslo, 0316Oslo, Norway.
[5] Currently at LOA, ENSTAParisTech, CNRS, Ecole Polytechnique, Universite´ Paris-Saclay,91762Palaiseau, France.

Generating low emittance beams from a plasma wakefield accelerator is an active area of research. Recently, several simulation papers have reported on the possibility of obtaining low-emittance electron beams using ionization injection [1-3]. Ionization injection refers to the insertion and acceleration of charge within an already formed wake by ionizing atoms or ions by the combined action of the electric field of the driver and the wake [4-7] or by using the electric field of additional sources such as one or more trigger laser pulses [1,3]. We have performed experiments to investigate the characteristics of injected electron beams that can be generated without using an external trigger such as a laser pulse. In particular, we will show that the injected beam, can be identified using an electron energy spectrometer, and that the energy spectra can be used to determine the charge, energy spread, and emittance of these beams. Additionally, we establish the scaling of the energy gain with length from 10 GeV to over 30 GeV by using a 30 cm and a 130 cm-long plasma source. The injected beams achieve energies that are up to 50% greater than the initial drive electron beam while maintaining a normalized emittance ~30 mm-mrad, which is over ten times smaller than that of the initial drive beam.

## 2. Experimental setup

The experiments described here were performed at the FACET facility at the SLAC National Accelerator Laboratory [8]. The plasma was formed by ionization of lithium vapor, and the ionization occurred either via the transverse electric field of the beam itself or by an auxiliary laser pulse. The lithium vapor was generated in a heat pipe oven [9], where liquid lithium was heated to temperatures of nearly 1000 °C. The resulting lithium vapor was contained by helium gas, which is in pressure balance with the lithium vapor as expected at thermodynamic equilibrium. The presence of helium as the buffer gas led to the creation of lithium density ramps. Where the density and pressure of lithium vapor falls, residual helium atoms maintain the pressure balance. The resulting number density of the neutral lithium and helium atoms is shown as an inset in figure 1. Two different configurations were used in this experiment: one was a 30 cm long column of $2.5 \times 10^{17}$ cm$^{-3}$ lithium vapor density and the other was a 130 cm column of $8 \times 10^{16}$ cm$^{-3}$ lithium vapor density.

The primary diagnostic used to observe the injected helium charge was an imaging electron energy spectrometer [10], as illustrated in figure 1. This diagnostic consisted of a dipole magnet, which was the dispersing element, in addition to two quadrupole magnets that were used to focus electrons at various energies. Toroidal charge monitors were used to observe excess charge generated during the plasma interaction [11], which is calculated as the difference between the upstream and downstream toroids.

Additionally, pulses from a 10 TW laser system were used to create a plasma [12] in order to study the effect of the preionization on the injection process as will be described in Section 5.

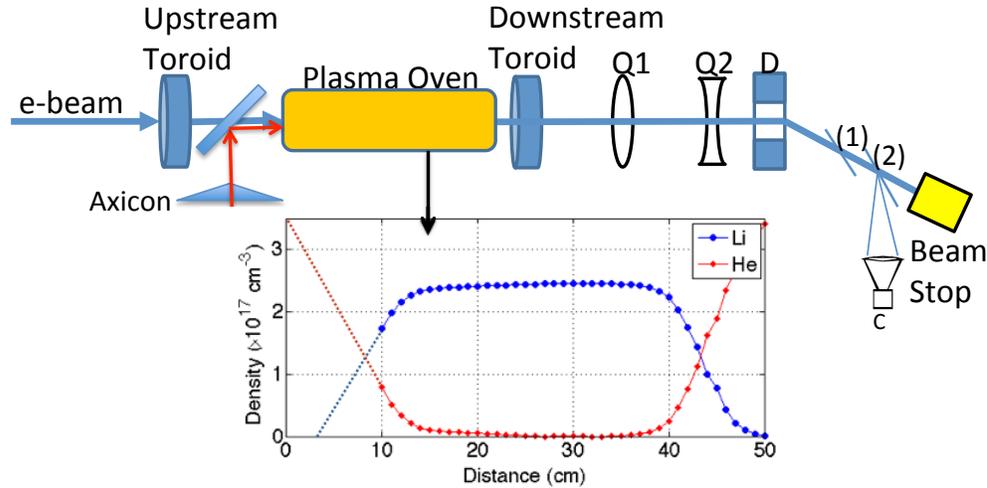

**Figure 1:** Experimental setup. Toroidal charge monitors upstream and downstream of the plasma monitor the input and output charge. The inset shows the neutral density profile of lithium and helium in a nominally 30 cm long oven. The spectrometer is formed by the dispersive element (D) and a pair of quadrupoles (Q1, and Q2) used to focus particular energies. Cherenkov radiation is produced by the dispersed electrons in the 5 cm air space between two Si wafers marked (1) and (2) and is recorded by the camera C. The electrons are dumped into the beam stop. An axicon optic was used to preionize the plasma (see section 5).

### 3. Observation of injected electrons

The first evidence for the injected charge comes from the difference in charge registered by the downstream and upstream toroids. This excess charge was first seen by E. Oz et al. [4] and subsequently analyzed by N. Kirby et al. [7] and N. Vafaei-Najafabadi et al. [11]. Although much of this charge comprises sub 100 MeV electrons from trapping occurring in the exit downramp, tens of pC charge with multi GeV energies was previously observed [7]. It was soon realized through theoretical analysis and computer simulations, that far from being a source of dark-current that would drain the energy stored in the wake, the ionization-injected charge could be a source of very low emittance electron beams. In this work we report our progress towards generating such beams. In these experiments as in the previous work, when the plasma was formed via beam ionization (without the presence of an ionizing laser), the electron spectra resulting from the interaction of the drive beam and the 30 cm plasma showed two distinct groups of electrons (see figure 2 for several examples). The first group of electrons, which belonged to the drive beam, can be observed on the spectra with energies spanning from 10 GeV to over 20 GeV. The second group, typically narrower than the first, is comprised of electrons between 6-10 GeV, which is completely disconnected from the first group in frames (a)-(c). This disconnect strongly suggests that the two groups had different origins and that the second group was produced in the course of the interaction of the drive beam with the plasma, meaning that this is the accelerated injected charge. The charge contained in the injected beamlets was between 10-20 pC. This charge estimate is based on the pixel counts on the spectrum, which is calibrated against the number of electrons in the drive beam in the case of no plasma interaction. The estimates for the beamlet charges are made for the cases where the charge of the injected electrons is clearly separate from those of the drive beam electrons. The peak energy of the injected charge ranged from 6-10 GeV, which implies a peak acceleration gradient of over 30 GV/m. The energy spread of the injected charge was several GeV, with the lowest observed value being 1.2 GeV FWHM for the case in figure 2b, or about 20%.

The divergence angle of the narrow injected beamlets was deduced by comparing the width of two features in the non-dispersive (horizontal) plane at energies different from the energy focused by the spectrometer. For data in figure 2, quadrupoles of the imaging spectrometer were set such that the electrons at 6.35 GeV were imaged. As both injected charge and the beam electrons can be observed at energies near 9 GeV in figure 2d, this image is particularly well suited for comparing the divergence of the two beams. The rms width of a slice of the injected charge at 8.4 GeV is 0.6 mm as compared to 2.8 mm for a slice of the drive beam at 9.4 GeV. The expected location of an electron can be calculated using transport matrix elements,
$\begin{bmatrix} x_1 \\ x_1' \end{bmatrix} = \begin{bmatrix} R_{11} & R_{12} \\ R_{21} & R_{22} \end{bmatrix} \begin{bmatrix} x_0 \\ x_0' \end{bmatrix}$, where $x_1$ and $x_1'$ are the transverse location and angle at the Cherenkov screen, $x_0$ and $x_0'$ are the transverse location and angle at the exit of the plasma, and $R_{mn}$ are the elements of the transport matrix, which are determined by our spectrometer configuration. In our system, $|R_{11}| < 2.5$, and $R_{12} = 10.9$ and $13.1$ for the 8.4 GeV (corresponding to injected-beam slice) and 9.4 GeV (corresponding to drive beam slice), respectively. Since the maximum value of $x_0$ is the radius of the blowout regime ($R_b \sim 26$ μm as will be shown in table 1), the value of $R_{11} x_0 \sim 70$ μm and therefore is negligible compared to the size of the above slices. This means that the difference in the size of the two beam slices is primarily due to divergence. Using the values of the beam rms and $R_{12}$, it is observed that the divergence of the drive beam is 3.8 times higher than the injected charge. For lineouts at other energies, this value remains between three and four. In the case of similar sizes for the drive beam and the injected beam at the exit of the plasma therefore, it can be concluded that the emittance of the injected charge is at least three to four times smaller than that of the drive beam. A more direct calculation of emittance using a high-resolution fluorescent screen is described in Section 6. Figure 2d also shows the difficulty of estimating the charge and energy spread in the cases where beamlets obtain higher energies. When the trapped beam and the drive beam electron signatures merge on the spectrometer, the charge and energy spread are difficult to estimate. It can be reasonably assumed that when the drive beam electrons experience a higher energy loss, there is a higher amount of charge that can be trapped because the larger fields accommodate more electrons before they are beam-loaded [13]. Therefore, the 10-20 pC represents a lower estimate of the injected charge.

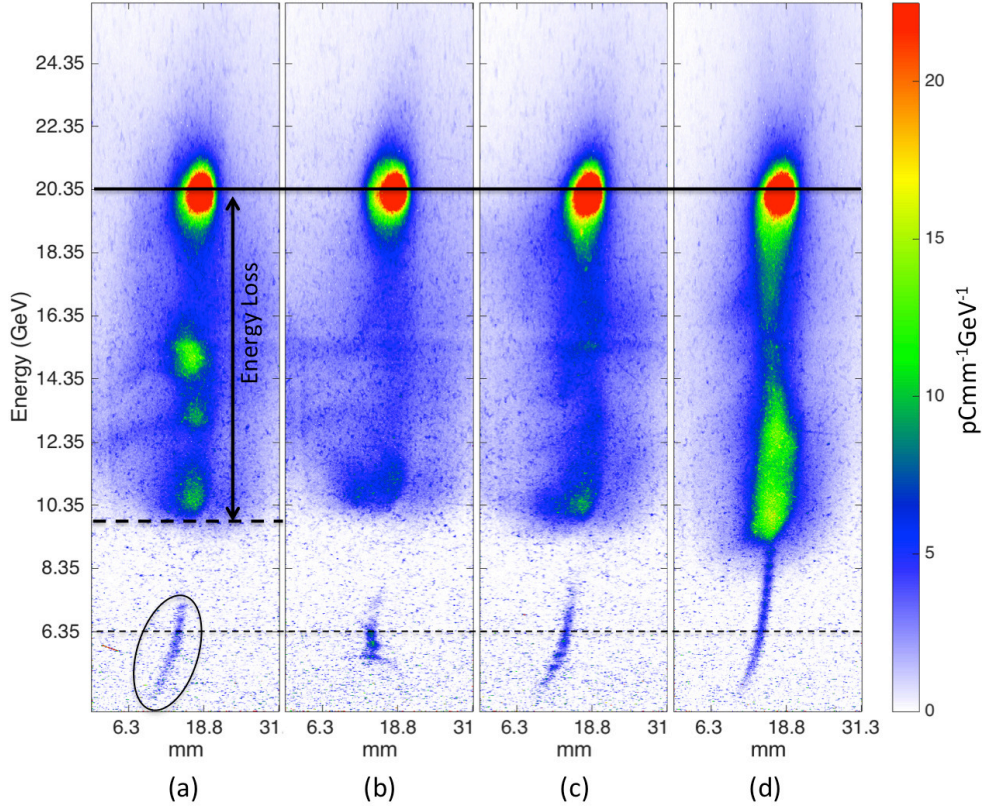

**Figure 2:** Energy spectra resulting from the interaction of the drive beam and the 30 cm plasma. The solid black line indicates the initial drive beam energy at 20.35 GeV. Electrons with energy of 6.35 GeV are focused by the spectrometer's quadrupole magnets, and marked by thin dashed black line. The injected charge in (a) is identified with an ellipse and the peak energy loss is indicated by a thick dashed black arrow. (b) The energy distribution of injected charge peaks at 5.8 GeV and has an energy spread of 1.2 GeV FWHM (20%). (d) Energy gain can reach over 10 GeV, in which case the maximum energy of the injected charge is difficult to discern as it merges into and overlaps with the energy loss feature of the drive beam.

### 4. Helium as source of injected charge (simulation)

The ionization of the first helium electron is identified as the source of the injected beamlet in a 2D cylindrically symmetric, particle-in- cell (PIC) OSIRIS simulation of the experiment [14]. An electron beam with $1.8 \times 10^{10}$ electrons is initialized with a beam size of $30 \times 30 \times 30$ μm$^3$. The moving window has a grid of 720x400 square cells, with the size of the cells being 0.05 $k_p^{-1}$, and all parameters normalized to $n_0 = 2.5 \times 10^{17}$ cm$^{-3}$. Ionization of helium and lithium are included in the code using the ADK tunnel ionization model, and each species has 16 particles per cells. Ionization fraction ($n_e/n_0$) is calculated by [15] using

$$W(s^{-1}) \approx 1.52 \times 10^{15} \frac{4^{n^*} \xi(eV)}{n^* \, \Gamma(2n^*)} \left(20.5 \frac{\xi^{3/2}(eV)}{E(GV/m)}\right)^{2n^*-1} \exp\left(-6.83 \frac{\xi^{3/2}(eV)}{E(GV/m)}\right), \tag{1}$$

$$\frac{n_e}{n_0} = 1 - \exp(-\int W dt). \quad (2)$$

Here, E is the electric field in GV/m, $\xi$ is the ionization potential in eV, $\Gamma$ is the mathematical gamma function, n* is the effective principal quantum number and is equal to $n^* \approx 3.69\, Z/\xi^{\frac{1}{2}}\,(eV)$, where Z=1 is the charge state of the ion. The density profiles of lithium and helium used in the simulation are similar to those shown in figure 1. The lithium density is modeled as a trapezoid with 10.4 cm long density ramps on either side of a 31.2 cm density plateau at to $n_0 = 2.5 \times 10^{17}$ cm$^{-3}$, and helium density is modeled as an inverted trapezoid, such that its density ramps down to zero at a point where the lithium plateau begins.

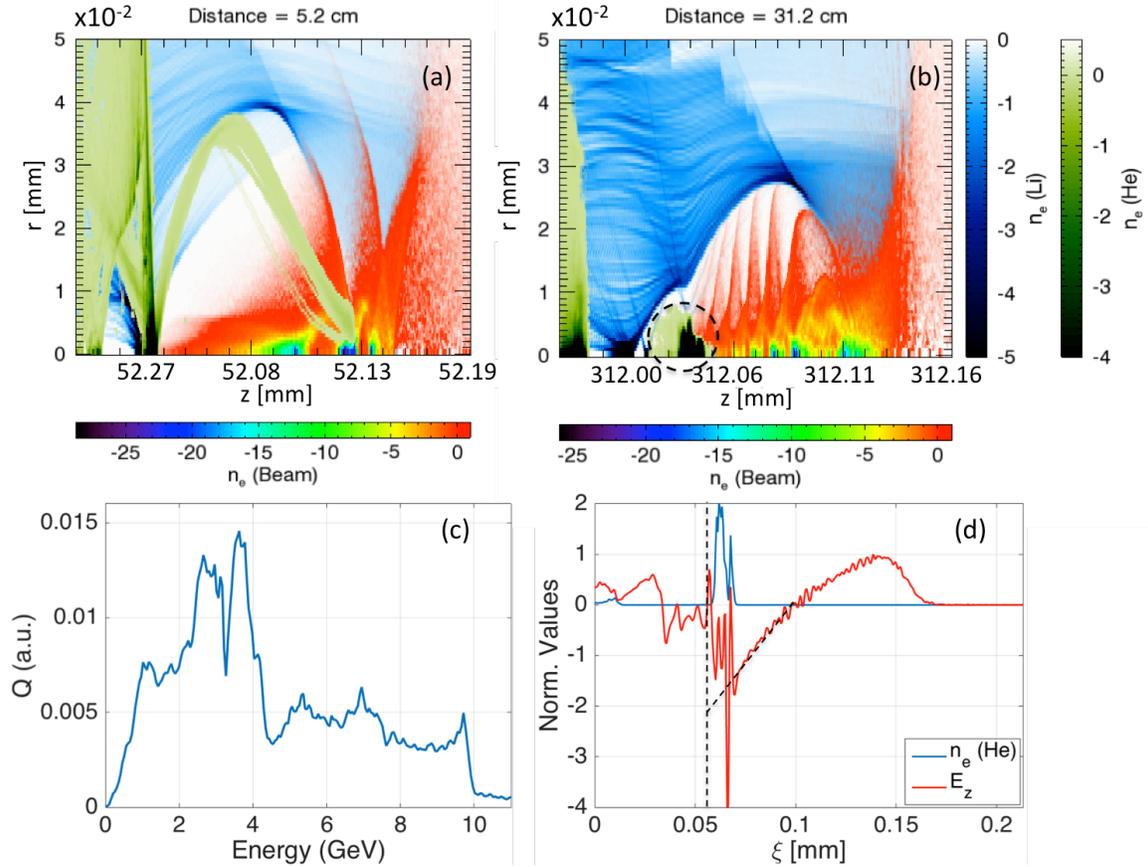

**Figure 3:** OSIRIS simulation results. (a) shows an r-z slice of the cylindrically symmetric simulation, where the e-beam has propagated 5.2 cm (halfway up the ramp). The color table for the drive beam electrons is shown below the figure. Lithium and helium electron densities are shown in blue and green, respectively. The process of helium electron injection can be observed. (b) r-z slice of simulation after the e-beam propagates 31.2 cm (20.8 cm in the plateau region). Color map for the beam electrons is displayed below the image. Lithium and helium electrons are shown with same color maps as (a). The Helium electrons at the back of the wake are encircled with a dashed black circle. (c) Energy spread of the helium electrons after the e-beam propagates 31.2 cm. (d) The on-axis longitudinal electric field (red) normalized to 36 GV/m and the on-axis density of the helium beam (blue) normalized to $6 \times 10^{19}$ cm$^{-3}$. The dashed black line indicates the useful accelerating field without beam loading.

Ionization of helium on the lithium density ramp occurs primarily due to the transverse electric field of the electron beam. The transverse spot size of the mismatched electron beam oscillates as the electron beam undergoes betatron oscillation [16,17]. Once the transverse size of the beam gets sufficiently small, it will lead to ionization of helium. The threshold transverse size that leads to appreciable ionization of helium can be calculated using the ADK tunneling model (equations (1) and (2)). The calculations show that a Gaussian electric field profile with $\sigma_z$=30 μm will need a peak field of 62.5 GV/m to ionize 10% of helium atoms. Given the equation of the transverse field of an electron beam, $E_r^{max} = 17.3\ GV/m\ (N/10^{10})(10\ \mu m/\sigma_r)(30\ \mu m/\sigma_z)$, where $N = 1.5 \times 10^{10}$ is the observed number of electrons participating in producing the wake in the simulation, the 10% ionization of helium requires a $\sigma_r^{th} = 4.1$ μm. During the propagation of the electron beam in the density ramp, the transverse size of various beam slices decreases to below this value, leading to ionization of helium. In figure 3a for instance, generation of helium electrons is clearly observed near the slice with $\sigma_r \sim$ 2 μm, generating over 100 GV/m of electric field, which easily ionizes helium. Although many of these electrons are left behind by the wake, some of these electrons are "trapped" and form a stable beamlet that is accelerated by the wake. This beamlet of electrons is shown to persist after 31.2 cm of propagation in figure 3b, with 20.8 cm of propagation occurring in the plateau region. The energy spectra of this beam after 31.2 cm propagation is shown in figure 3c, and it has several features that corroborate those of the beamlets observed in the experiment such as a wide energy spread and a peak energy of nearly 10 GeV. It is also interesting to note that a small number of electrons in the simulation gain as much as 20 GeV of energy. The on-axis electric field and density of the trapped charge are shown in figure 3d. The field can be observed to be heavily beam loaded, with the peak loaded gradient of the wake around a normalized value of -1 or 36 GV/m, which is half the value of the useful accelerating field, defined by linearly extending the field to the back of the wake (dashed black lines in figure 3d).

**5. Helium as the source of the injected charge (experiment)**

The trapping condition for an electron in PWFA is expressed as $\Psi_f - \Psi_i < -1$ [4,5]. Here, $\Psi$ is a normalized pseudo potential and is equal to $\Psi = (\phi - A_z)/(mc^2)$, where $\phi$ and $A_z$ are the scalar potential and the z component of the vector potential, respectively. Here, $\Psi_i$ refers to the value of the pseudo potential where the electron is born and $\Psi_f$ is the final value of this potential as an electron slips back in the wake. If $\Psi_f - \Psi_i < -1$ for these electrons, then they have achieved the wake's velocity and therefore are trapped and this tends to occur near the back of the region of blowout. For a wake in the blowout regime, electrons which are born via laser preionization of both lithium atoms and the buffer helium atoms will have a $\Psi_i = 0$ resulting in $\Psi_f - \Psi_i > -1$ [5] leading to no trapped electrons. For the case of no laser preionization (most of the results in this paper), the lithium atoms are easily ionized by the electron beam itself and are thus necessarily born near the head of the bunch, just prior to their radial blowout. As in the laser trigger case, since they are born *before* the wake formation, their $\Psi_i$ will also be near zero. However, as the beam within the lithium ion column of the wake executes betatron oscillations, its radial field at a local minimum of its spot size can be sufficient to ionize the (formally neutral) helium atoms *within* the wake, thus reducing $\Psi_i$. If the wake is sufficiently large, these helium electrons will become trapped towards the rear of the wake.

Therefore, the origin of the low divergence electron feature seen in figure 2 as being ionization injection of helium electrons can be experimentally tested using an intense laser pulse to preionize the helium atoms in the density up-ramp of the lithium source before the electron bunch arrives to produce the wake. A 500 mJ, 100 fs laser pulse from the nominally 10 TW FACET Ti:Sapphire laser was focused using an axicon optic with an angle of $\alpha = 1.5$ degrees. The axicon was located 95 cm upstream of the start of the lithium density plateau. The laser intensity distribution after the axicon for a flat-top laser can be calculated as [18]

$$I(r,z) = I_0(2\pi)^2 \frac{z\delta \sin(\delta)}{\lambda \cos^2(\delta)} J_0^2(k\delta r), \tag{3}$$

where $k = 2\pi/\lambda$, with lambda the laser wavelength, and $\delta$ is the angle of approach as calculated from $sin(\delta + \alpha) = n \sin(\alpha)$. The axicons used at FACET are made from fused silica, where n = 1.45. The laser pulse has an approximately flat-top profile with a width of 40 mm at the location of the axicon, leading to intensity of $I_0 = 4.0\times10^{11}$ Wcm$^{-2}$ on the axicon. According to equation (3), a flattop laser profile produces a linearly increasing intensity on axis downstream of the axicon as shown in figure 4a. Also shown in this figure are the density profiles of the helium and the lithium. Using this intensity distribution and modeling the ionization of helium using the ADK tunnel ionization rate [15], we can calculate the ionization map for the helium atoms in the interaction region using equations (1) and (2). The resulting ionization map (figure 4b) shows that the helium electrons are fully ionized at the location of the helium density ramp. If the helium atoms are ionized before the arrival of the electron beam, they are blown away by the head of the electron beam just like the lithium electrons and therefore no injection into the wake can happen. We would therefore expect the additional beamlets seen in figure 2 to disappear. Indeed, when laser pulses were sent ahead of the electron beam in this experiment, the trapped charge disappeared in almost every case. An example of such side-by-side comparison using the short plasma can be observed in figure 5a and figure 5b, where the difference in the two shots is the presence of the laser. Similarly, the low-divergence, high-energy signal seen with the longer length plasma (see Section 6) disappears when the laser pulse is sent ahead of the electron beam as seen in figure 5c and figure 5d. In the case of the long plasma, a 0.75-degree axicon was used 120 cm upstream of the lithium plateau with the other laser parameters unchanged. Although similar calculations as presented in Figure 4b indicate possible 50% ionization of the helium, the

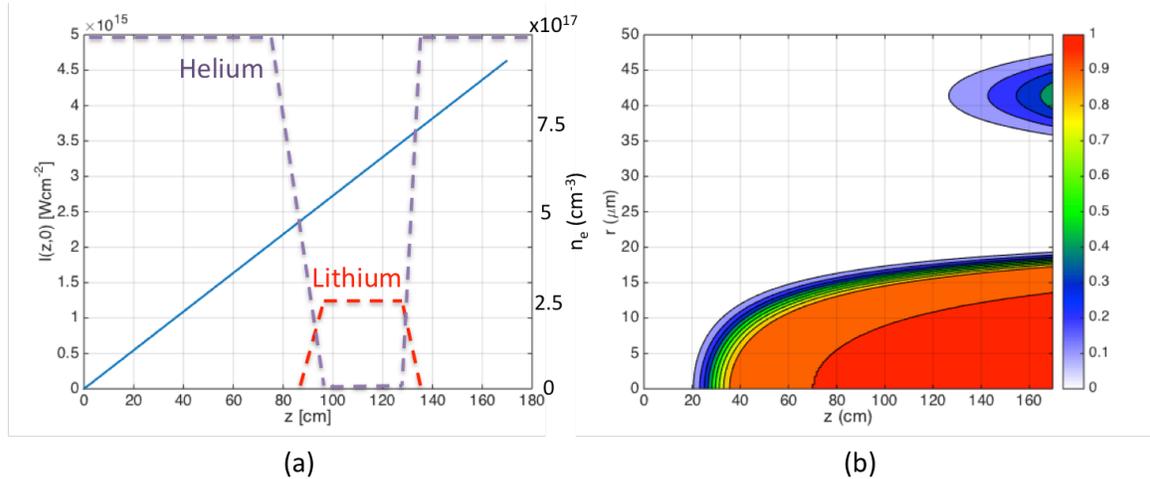

**Figure 4:** Calculation of the region in space where the laser is intense enough to ionize helium atoms. (a) Blue solid line is the intensity profile of a 500 mJ, 100 fs laser pulse with a 40 mm wide flat top profile incident upon a 1.5 degrees axicon. The neutral density profiles of lithium and helium are overlaid on the intensity profile as dashed red and purple lines, respectively. (b)

Ionization fraction contour ($n_e/n_0$) of helium subject to laser intensity in (a) calculated from the integrated ionization rate (equations (1) and (2)) over 100 fs.

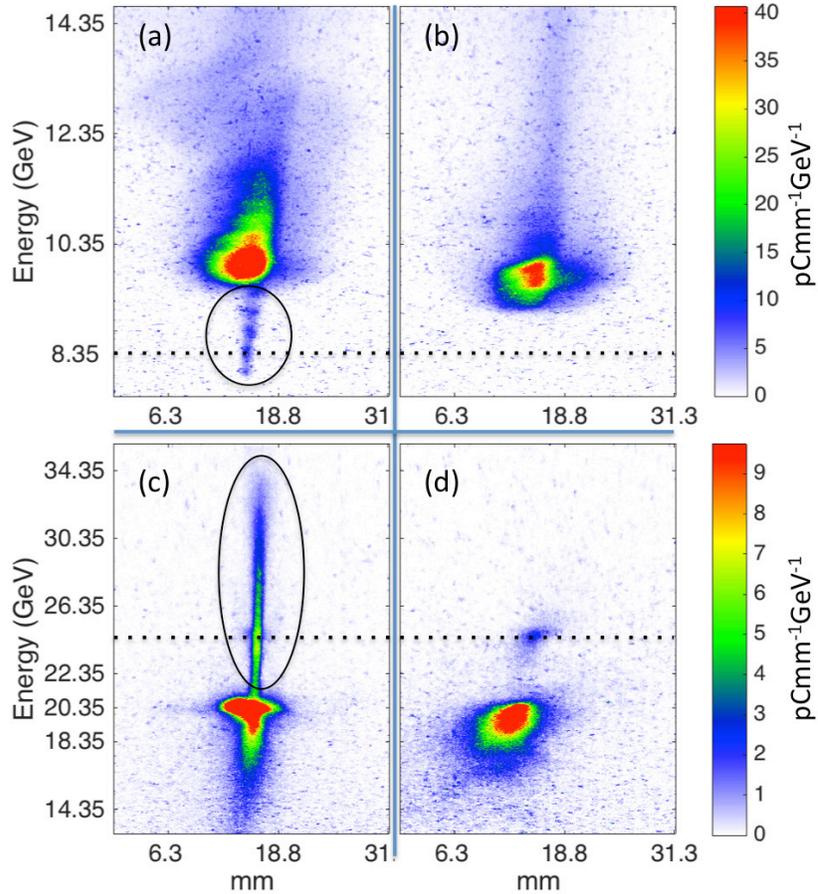

**Figure 5:** Spectra resulting from the interaction of the electron beam with (a) the 30 cm lithium vapor without preionization laser. (b) 30 cm lithium vapor with laser pulse preionizing the plasma. (c) 130 cm lithium vapor without preionization laser. (d) 130 cm lithium vapor with laser pulse preionizing the plasma. The energy focused by the quadrupoles of the spectrometer is marked with a dashed black line. The color table for (a) and (b) is defined by the top color bar and the one for (c) and (d) is defined by the bottom color bar. Injected charge is marked with solid black ellipses.

presence of the laser nevertheless removed the injected beam. Furthermore, when the laser energy was reduced to 1/4 its nominal full energy, the injected charge reappeared. In other words, as long as the laser energy was above the threshold of ionization of lithium, but not helium, the injected charge features appeared in the experiment strongly indicating that the injected charge comes from the ionization of the helium atoms by the combined action of the beam and the wakefields within the fully formed wake.

## 6. Plasma length scaling

The injected beamlets were also observed when a second, longer plasma source was used. The plasma density was reduced from $2.5\times10^{17}$ cm$^{-3}$ in the case of short plasma to $8\times10^{16}$ cm$^{-3}$ in the case of the long plasma to increase the plasma wavelength and thus reduce the number of accelerated drive beam electrons. This was done to enhance the visibility of the injected charge on the spectrometer. Because of the high energy of the drive beam ($\gamma \sim 40{,}000$), each longitudinal slice of the electron beam maintains its relative distance with respect to the other slices during the interaction and does not dephase with respect to the wake. Therefore, with increasing plasma length, we expect an increase in the energy of the injected charge for as long as the beam can drive the wake. The expected scaling of the electron energy can be estimated using the equations for the accelerating fields in the blowout regime [19]. The peak accelerating field is given by $eE_{max}/(mc\omega_p) = -\frac{1}{2}k_p R_b$, where $R_b$ is the maximum blowout radius and $k_p = \omega_p/c$ is the inverse skin depth of the plasma. In the case of a bi-Gaussian beam, $k_p\sigma_z \approx 1$, $k_p\sigma_r \ll 1$, and $k_p R_b \approx 2\sqrt{\Lambda}$, where $\Lambda$ is charge per unit length and can be expressed as $\Lambda = n_b/n_0 (k_p\sigma_r)^2$, where $n_b$ is the beam density. Thus in the blowout regime, we can make a rough estimate for the expected energy from the short and the long plasma source as shown in table 1.

**Table 1:** The expected interaction parameters in the short and the long oven. L is the length of the oven as measured using the FWHM of the density profile. All other symbols are defined in the text.

|  | Long Plasma 130 cm | Short Plasma 30 cm |
|---|---|---|
| $n_0$ (cm$^{-3}$) | $8\times10^{16}$ | $2.5\times10^{17}$ |
| $n_b$ (cm$^{-3}$) | $4.7\times10^{16}$ | $4.7\times10^{16}$ |
| $\Lambda$ | 1.49 | 1.49 |
| $k_p R_b$ | 2.44 | 2.44 |
| $R_b$ (µm) | 46 | 26 |
| $E_{max}$ (GV/m) | 33.3 | 58.8 |
| L (m) | 1.3 | 0.3 |
| $\Delta W_{max}$ (GeV) | 43 | 18 |

In the case of the short plasma, the peak energy gain ($\Delta W_{max}$) of the helium electrons was observed to be in the 8-10 GeV range. One possibility suggested by simulation (figure 3d) for the difference between this range of numbers and the 18 GeV energy gain that was calculated from the nonlinear wakefield theory is that the wake is beam loaded to about half of its unloaded strength.

For the long plasma, the theoretical peak unloaded energy gain is 43 GeV (table 1). Assuming a similar level of beam loading for the plasma wakefield in the long oven as in the case of short oven, the expected value for the energy gain in the long oven is in the range of 19-24 GeV. The results observed from the long plasma corroborate this estimation of energy gain as seen in figure 5c and figure 6. In some instances, such as figure 6a, a disconnect can be observed between the electrons from the main beam and the trapped injected electrons, as was also observed in the short-plasma results. The reason for this is the difference in the relatively larger divergence angle of the drive beam electrons compared with the smaller divergence angle of the injected charge. Consequently, whereas the trapped electrons remain visible well away from the spectrometer's energy focus, accelerated electrons from the tail of the drive-beam have their maximum visibility only near the spectrometer focus (23.35 GeV) as shown in figure 6b and are progressively defocused above and below this point. Meanwhile, the discrete injected beam (e.g. figure 6b) appears as a long narrow stripe even though it is ~7 GeV away from the spectrometer focus. In some of the data, the injected helium bunch is well aligned with the signature of the accelerated particles from the drive beam (figure 6c). It is important to note that comparable narrow divergence features do not appear in the case of the short plasma at energies over 20 GeV, supporting the idea that these narrow-energy-spread beams have a different source than the drive beam.

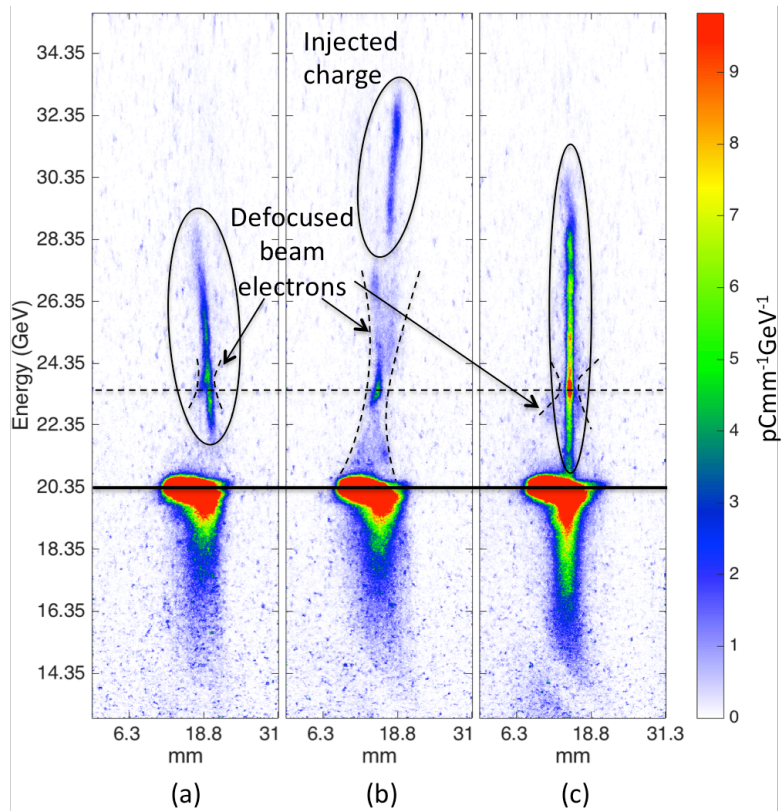

**Figure 6:** Spectra resulting from the interaction of the drive beam with the 130 cm long, $8 \times 10^{16}$ $cm^{-3}$ plasma. The energy focused by the spectrometer's quadrupole magnets is marked by the dashed black line. The solid black line represents the initial energy of the drive beam. The injected beam in each frame is marked with a solid black ellipse. The outer edges of the defocused beam electrons are traced with two dashed black curves. (a) Example of a narrow divergence feature disconnected from drive beam—the charge is 25 pC. (b) Highest observed energy of the injected charge with limited energy spread. (c) Example of a low divergence injected beam overlapping with the high-energy-spread accelerated drive beam electrons.

The emittance of the injected beam can be calculated using spectra recorded on a high resolution LANEX screen (instead of the Cherenkov detector shown in figure 1). An example of data from this screen is shown in the inset in figure 7a. On this high-resolution screen, horizontal lineouts at various energies close to the imaging energy of the spectrometer are fitted with a Gaussian curve and the rms value of the Gaussian at each energy is measured, resulting in a $\sigma_x(E)$ curve, an example of which is plotted in figure 7a. An iterative, parametric fit based on the configuration of the spectrometer is made to the resulting curve of $\sigma_x(E)$. Using this fit, the beam parameters such as normalized emittance and the location of the waist can be calculated such as was done in [20]. A histogram of emittance values for 65 shots belonging to one dataset is displayed in figure 7b, showing emittance values as low as 14 mm-mrad with a mean value of 32 mm-mrad. However, because of the low value of the emittance, only a small number of pixels (on the order of ten pixels) contribute to the measurement. Therefore, there is an uncertainty in the value of the emittance as the measured rms values approach the point-spread function of the optical system. This function is not known for this system, but its effect on the measured values of emittance can be examined by assuming various values for the point-spread function. For instance, the measured value of 21 mm-mrad represents a beam with emittance of 19 and 15 mm-mrad if the point spread function has values of 4 and 7 pixels, respectively. In other words, values reported in figure 7b represent upper estimates for the emittance of the injected beam in the experiment. It is clear that the emittance of the injected electrons (measured in the non-dispersive, x plane) is at least a factor of 10 smaller than that of the drive beam ($\epsilon_{n,x} \sim 358$ mm-mrad) when it is injected into the plasma [21]). The difference in emittance between the injected beam and the drive beam shows the potential for the ionization injection to produce a beam that has a higher brightness than the initial drive beam if the energy spread of the injected beam can be reduced while maintaining its high current.

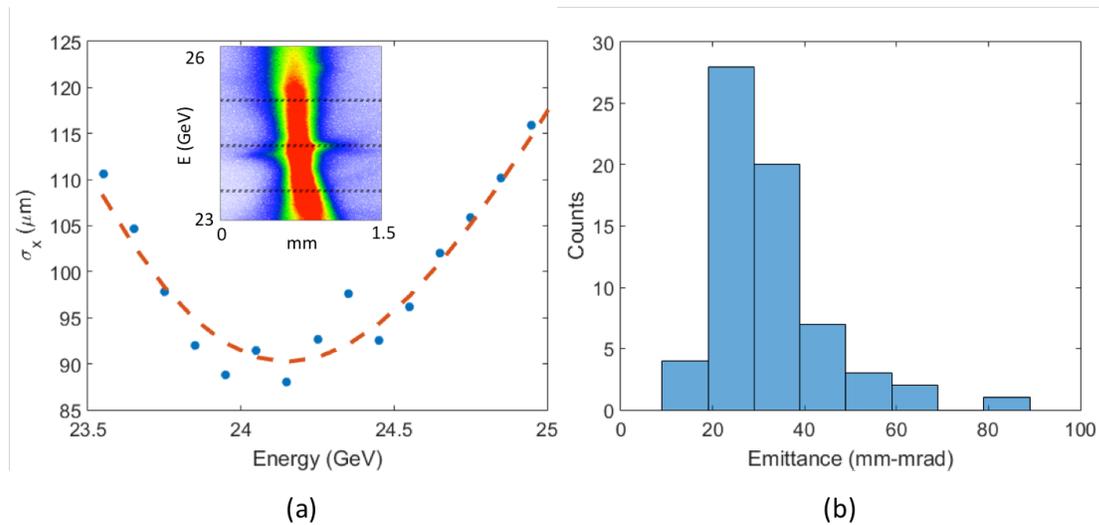

**Figure 7:** Emittance measurement for the injected bunch in the long plasma. (a) Blue circles indicate the values of $\sigma_x$ at different values of energy for a single spectrum image, which is shown in the inset. The top and bottom black dashed lines in the inset correspond to the highest and lowest energies for which $\sigma_x$ is calculated. Red dashed line represents the best fit to the data, resulting in emittance value of 36 mm-mrad. (b) Histogram of emittance values ($\epsilon_{n,x}$) for a dataset, where 65 data points belonging to the same dataset could be fit and measured.

## 7. Conclusions

We have shown the ability of ionization injection to produce electron beams at a gradient of ∼ 40 GeV/m and with narrow divergence. The electrons are identified as originating from a different source because they are separated from the main drive bunch on the spectrometer and in addition have at least a factor of 10 smaller divergence and emittance than that of the drive bunch in the plane of the measurement. The injected beam has a charge ranging in tens of pC and has energy spread of several GeV. In order to reduce the energy spread and emittance of the injected charge, the injection region should be limited such that the injection process occurs over a shorter length. Such confined region of helium impurity can be obtained using a capillary setup to inject helium gas within a much longer column of hydrogen and thereby create a localized gas jet used as a narrow region for injection. This technique may lead to the generation of super bright beams needed for future applications.

## Acknowledgements

Work at UCLA was supported by DOE grant number. DE-SC0010064 and the NSF grant number PHY-1415386. Work at SLAC was supported by DOE contract number DE-AC02-76SF00515. Simulations used the Hoffman cluster at UCLA.